\documentclass[aps,prl,twocolumn,groupedaddress,showpacs]{revtex4}

\usepackage{graphicx}

\begin{document}

\title{Feeding upon negative entropy in a thermal-equilibrium environment}

\author{Bruno Crosignani}
\affiliation{Department of Applied Physics, California Institute of Technology, Pasadena,
California 91125, USA and \\ Dipartimento di Fisica, Universita' dell'Aquila, 67010 L'Aquila, Italy and \\
Istituto Nazionale per la Fisica della Materia, Universita' di Roma "La Sapienza", 00185 Roma, Italy}

\author{Paolo Di Porto}
\affiliation{Dipartimento di Fisica, Universita' dell'Aquila, 67010 L'Aquila, Italy and \\
Istituto Nazionale per la Fisica della Materia, Universita' di Roma "La Sapienza", 00185 Roma, Italy}

\author{Claudio Conti}
\affiliation{Istituto Nazionale per la Fisica della Materia, Universita' Roma Tre, 00146 Roma, Italy}

\date{\today}

\begin{abstract}

The validity of the Second Law of thermodynamics, indisputable in the macroscopic world, 
is challenged at the mesoscopic level: 
a mesoscopic isolated system, possessing spatial dimensions of the order of a few microns, 
is capable, as shown by a straightforward kinetic analysis, to exhibit a perpetuum mobile 
behavior associated with large negative variations of the Clausius entropy of the system. 
This violation of the Second Law is expedient for devising a cyclic process 
through which an isolated system can extract energy from a surrounding thermal bath.
\end{abstract}

\maketitle

\begin{figure}
\centerline{\includegraphics[width=0.35\textwidth]{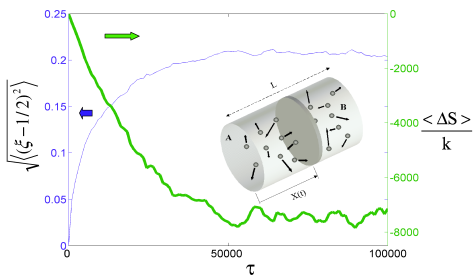}}
\caption{\label{f1} (Colors online) The adiabatic piston: an
insulating cylinder divided into two regions by a movable,
frictionless and insulating piston. Left (thin line): evolution as
a function of $\tau = t/t_0$ of the rms deviation of the piston
position from its initial value $\xi(0)=1/2$, for $\mu =0.5$ and
$N = 3 \cdot 10^4$ (over 1000 realizations). Right (thick line):
time evolution of the entropy. }
\end{figure}

As first clearly  stated by Schroedinger ``a living  organism tends to
approach the  dangerous state of  maximum entropy, which is  death. It
can only keep  aloof from it, i.e., alive,  by continually drawing from
its  environment  negative  entropy.  What  an organism  fed  upon  is
negative entropy".\cite{Schroedinger}  Is it possible  for an organism
immersed  in  a   thermal  bath  to  be  insulated,   and  thus  avoid
thermalization,   and   still  be   able   to   reduce  its   entropy?
Unfortunately,  the  above  scenario   contradicts  one  of  the  most
cherished laws of  physics, that is the Second  Law of Thermodynamics,
which has  victoriously resisted all  the attempts aimed at  finding a
particular  case where  it could  be  violated.  However,  one has  to
consider that  the Second  Law has been  formulated in the  context of
macroscopic  physics  and it  is  in this  context  that  it has  been
successfully  applied  and  verified.   While the  Second  Law,  being
inherently statistical in its nature, cannot carry over to microscopic
cases where the number of  involved particles is too small, its limits
of  validity  are not  well  understood  in  the region  bridging  the
macroscopic to the microscopic  world. In particular, the intermediate
mesoscopic  regime, which includes  part of  the biological  realm, is
still \textit{terra incognita} and  open to  possible surprises.  Recent years
have actually  witnessed a growing  interest in the  thermodynamics of
small-scale non-equilibrium devices, especially in connection with the
operation and efficiency of  Brownian motors.\cite{P1} We believe that
we have  been able to  devise a particular  case in which  an isolated
object, placed  in a thermal-equilibrium  environment, is nevertheless
capable of reducing its entropy or, equivalently, to extract work from
the thermal bath surrounding it. We have identified this object with a
mesoscopic system, typically of the  size of a few microns, consisting
of two  isolated cavities, filled  with a common number  of molecules,
separated   by  a   movable  adiabatic   impermeable   partition  (see
Fig.\ref{f1}). Its macroscopic equivalent  is the so-called adiabatic piston,
sometimes referred  to as ``enigmatic''  because the description  of its
dynamical evolution toward equilibrium  is far from being trivial, the
problem  being  actually  undetermined  from  the  point  of  view  of
elementary thermodynamics.\cite{P3} A careful analysis of the problem, which
has been  tackled a few years  ago by an  approach based on  gas kinetic
theory,\cite{P4} shows  that thermodynamic fluctuations  cannot be neglected
at the  mesoscopic level and that  they actually play a  major role in
determining the dynamic evolution of this isolated system. As shown in
this paper, its behavior implies  that the system fails to settle down
to  a well-defined  equilibrium state  and undergoes  negative entropy
fluctuations whose  ensemble average over many replicas  of the system
can be quite large.  Thus, on  the average, the system is able to feed
on negative  entropy. It is also  possible to show how  the system can
function  as  an  engine  and  turn the  entropy  extracted  from  the
surrounding  thermal  bath into  work.  These  remarkable results  are
derived in  the frame  of a model  which does  not contain any  ad hoc
parameters and appears to  possess some simple universal feature which
makes  it a  potential  candidate for  supporting  some still  unknown
ubiquitous biological process.

Our system consists of two isolated cavities A and B, filled with a common number of moles of the same gas,
separated by a movable impermeable partition which is assumed to be initially at rest in a position
corresponding to equal volumes and temperatures (and thus pressures) in the two cavities. Under these
conditions, one would expect the wall to remain at rest, apart from microscopic random oscillations around
the equilibrium position.

This is obviously the case in most ordinary situations, where the
wall thermal conductivity is large enough to maintain a common
temperature in the two cavities, so that any small random
displacement of the wall is immediately reversed by the induced
pressure change. Conversely, if thermal conductivity is
negligible, it will be shown below that the wall can actually
undergo sizeable random displacements, during which the pressures
on the two sides remain the same. In each realization, the system
evolves and the total entropy variation, from the initial state in
which both cavities have the same volume $V_0 / 2$ and temperature
$T_0$ to a generic one $(V_A,T_A ; V_B,T_B)$, is
\begin{eqnarray} \label{eq1}
\Delta S &=& \Delta S_A + \Delta S_B = n c_p \ln \frac{T_A}{T_0} + n c_p \ln \frac{T_B}{T_0} \nonumber \\
         &=& n c_p \ln \left[ 1 - \left( \frac{\Delta V}{V_0 /2} \right)^2 \right],
\end{eqnarray}
where $\Delta V = V_A - V_0 / 2$, $c_p$ being the molar heat at constant pressure and $n$ the common number
of gas moles on each side. The {\it ensemble average} $\Delta S$ of this intrinsically negative entropy
change can assume large values in specific mesoscopic situations, as it will be proved in this paper. Note
that in deriving Eq.(\ref{eq1}), we have exploited the relation $T_A + T_B = 2 T_0$, which corresponds to
neglecting the kinetic energy of the wall. In fact, under the adiabatic hypothesis, it can be regarded as a
body possessing a single (translational) degree of freedom. As a consequence, its energy and entropy are of
the order $k T_0 / 2$ and $k$, respectively, and can be neglected.

In order to corroborate the above considerations, we investigate the so-called ``adiabatic-piston problem'',
\cite{P3,P4} dealing with a system which, in the language of classical thermodynamics, consists of an
isolated cylinder divided into two parts by means of a frictionless adiabatic piston, each section containing
the same number n of moles of the same perfect gas (see Fig.(\ref{f1})).
Initially, the piston is held in the fixed position corresponding to equal volumes and pressures
in the two sections. Once the piston is released, the system evolves in a labile quasi-equilibrium state,
driven by the random motion of the wall induced by the elastic molecular collisions, while the gas pressure
on both sides remains equal to the initial one. \cite{P4,P5} We describe the dynamics of the system in terms
of the random variable $X(t)$, that is the instantaneous position of the piston, whose ensemble-average value
$\langle X(t) \rangle$ remains equal to $L/2$ for obvious symmetry reasons. The normalized mean-square value
$\langle[X(t)-L/2]^2\rangle /(L/2)^2 $ represents the ensemble-average of the quantity $(\Delta
V)^2/(V_0/2)^2$ appearing in Eq.(1) and determines the corresponding entropy decrease undergone by the
system.

The system evolution from an
initial configuration of macroscopic equilibrium, that is {\it
equal} pressures and temperatures on both sides and piston at
rest, is described by the
stochastic equation\cite{P5}
\begin{eqnarray} \label{eq3}
\frac{d^2 X}{d t^2} &+& \sqrt{\frac{16 N k T_0}{\pi \mu M L}}
                    \left( \frac{1}{\sqrt{X}} + \frac{1}{\sqrt{L-X}} \right) \frac{d X}{d t} \nonumber \\
                    &+& \frac{2}{\mu} \left[ \frac{X-L/2}{X(L-X)} \right] \left(\frac{d X}{d t}\right)^2= a(t),
\end{eqnarray}
where $k$ is Boltzmann's constant, $N$ the common number of
molecules on each side, and $\mu = M/M_g$, $M$ and $M_g$ being
respectively the mass of the piston and the common value of the
gas mass in each side. Determining
the correct expression of $a(t)$ is a delicate task, since the
standard Langevin approach does not in general carry over to
nonlinear dynamical systems \cite{P6}, as the one described by
Eq.(\ref{eq3}). In order to take advantage of the Langevin method,
we linearize the above equation: a) by considering small
displacements around the starting position, that is
$|(X-L/2)|/(L/2) \ll 1$, and b) by approximating the square of the
piston velocity $(dX/dt)^2$ with its mean-square velocity $k
T_0/M$ (both hypotheses will be proved consistent {\it a
posteriori}). Proceeding in this way, a straightforward
application of the dissipation-fluctuation theorem yields $\langle
a(t) a(t') \rangle = [8(2 m k T/ \pi)^{1/2} P S/M^2]
\delta(t-t')$, where $m$ is the mass of the individual molecule,
$P$ the common pressure on the two sides of the piston and $S$ its
area. After introducing the variable $x=X-L/2$, the linearized
form of Eq.(\ref{eq3}) reads
\begin{eqnarray} \label{eq4}
\frac{d^2 x}{d t^2} &+& 8 \sqrt{\frac{2 N k T_0}{\pi \mu M L^2}} \frac{d x}{d t}
                     + \frac{8 k T_0}{\mu M L^2} x = a(t).
\end{eqnarray}
We now observe that the above equation is formally identical to the one describing the Brownian motion of a
harmonically-bound particle of mass $M$, that is
\begin{equation} \label{eq5}
\ddot{x} + \beta \dot{x} + \omega^2 x = A(t),
\end{equation}
where $A(t)$ is the Langevin acceleration, a problem which has been thoroughly described in the literature
\cite{P7}. By comparing Eqs.(\ref{eq4}) and (\ref{eq5}), we can obviously apply the results of \cite{P7} to
our case by identifying $\beta$ with $8(2N k T_0 / \pi \mu M L^2)^{1/2}$ and $\omega^2$ with $8 kT_0 / \mu M
L^2$. Whenever $\beta \gg \omega$ (``overdamped'' case), which in our situation is equivalent to the
obviously satisfied relation $\sqrt{N} \gg 1$, the analysis carried out in \cite{P7} naturally highlights the
existence of two significant time scales $t_{th}=1/2\beta$ and $t_{as}=\beta/2 \omega^2$ ($t_{as} \gg
t_{th}$). They represent the thermalization time $t_{th}$, i.e., the time over which the mean-square velocity
$\langle (dx/dt)^2 \rangle$ attains its equipartition value $kT_0/M$, and the much longer time $t_{as}$ over
which the mean-square displacement reaches its asymptotic value $\langle x^2 \rangle = k T_0 /M \omega^2$.
For our system $t_{as} = (NL/w)(M/M_g)/ \pi^{1/2}$ and $t_{as} / t_{th} = 16 N / \pi$, where
$w=(2kT_0/m)^{1/2}$ is the most probable velocity of the gas Maxwellian distribution function. Therefore,
since $N \gg 1$, $t_{th}$ is much smaller than $t_{as}$, a circumstance which justifies {\it a posteriori}
the replacement of $(dX/dt)^2$ in Eq.(\ref{eq3}) by its average value $k T_0/M$.
The asymptotic value of the mean-square displacement of the piston from
its central position reads
\begin{equation}
<x^2>=\frac{k\,T_0}{m\omega^2}=\frac{\mu}{2}(\frac{L}{2})^2\;\textit{,}
\end{equation}
so that, in the limit $\mu \ll 1$ (that is, small piston mass
with respect to gas mass), the above assumptions a) and b) allowing us to linearize Eq.(\ref{eq3}) are both
satisfied.

We can now evaluate the entropy change $\langle \Delta S \rangle$ averaged over many realizations of our
system, starting from the initial state corresponding to $X(t=0)=L/2$. Since in our case $\langle (\Delta
V)^2 \rangle /(V_0/2)^2 = \langle x^2 \rangle /(L/2)^2 = \mu/2 \ll 1$, Eq.(\ref{eq1}) approximately yields
\begin{equation} \label{eq6}
\frac{\langle \Delta S \rangle}{k} = - \frac{c_p}{2R} N \mu,
\end{equation}
which corresponds to a {\it large entropy decrease} whenever $\mu
N \gg 1$. The question naturally arises: what are the spatial and
temporal scales over which this {\it violation of the second law
of thermodynamics} can actually occur? Before answering this
question, the conceptual meaning of the result implied by
Eq.(\ref{eq6}) has to be clarified. To this aim, we note that,
according to standard thermodynamics, {\it any closed system in an
equilibrium state, once an internal constraint is removed,
eventually reaches a new equilibrium state characterized by a
larger value of the entropy}. The case considered in this paper
represents a remarkable exception to this statement. More
precisely, our closed system is made up of the two gases and the
initial internal constraint is provided by the piston held in the
central position. After the piston is released, our analysis shows
that {\it no final equilibrium state is eventually reached}, since
the piston keeps wandering, performing random oscillations around
X=L/2, with mean-square amplitude $\langle x^2 \rangle = (L/2)^2
\mu/2$.

\begin{figure}
\includegraphics[width=0.35\textwidth]{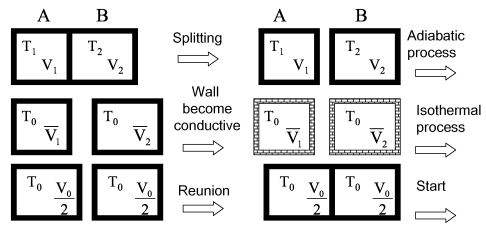}
\caption{Cyclic process through which work can be extracted from a
thermal bath at temperature $T_0$. } \label{fsplitting}
\end{figure}

The system behaves as a \textit{perpetuum mobile} of the second kind.
If embedded in a thermal bath at temperature $T_0$ and pressure
$P_0$, one can devise a process through which work can be extracted
from this environment. To this end, let us assume our system to be
capable of splitting into the two separate cavities $A$ and $B$ if
a large fluctuation $\Delta\,V$ occurs (see Fig.
\ref{fsplitting}). The two cavities can then undergo a reversible
adiabatic process which drives them back to the initial
temperature $T_0$. If, at this point, the wall of the cavities
becomes thermally conductive for a finite amount of time, the two
cavities can be brought back to the initial volume $V_0/2$ through
a reversible isothermal process at the bath temperature $T_0$. Let
us now assume that the wall recovers its insulating nature and
that the two cavities are reunited again: the total work extracted
in this cyclic process is precisely $W=T_0|\Delta\,S|\cong\,k\,T_0\,N\mu$, where
$\Delta\,S$ is the negative entropy variation associated with
$\Delta\,V$.

We now return to the question of the spatial and temporal scales
over which our results apply. To this end, let us consider the
specific case of a gas under standard conditions of temperature
and pressure, for which, expressing hereafter $L$ in microns, $N
\cong 3 \cdot 10^7 L^3$. According to Eq.(\ref{eq6}), if we refer
as an example to a biatomic gas, we have $\langle \Delta S \rangle
/k \cong -5 \cdot 10^7 L^3 \mu$ and, recalling the expression of
$t_{as}$, that is $t_{as}=NL\mu/w\sqrt{\pi}$, we obtain $t_{as} \cong 5
\cdot 10^{-2} \mu L^4 \,sec$ (having assumed $w \cong 4 \cdot
10^8$ microns/sec, molecular oxygen). This {\it extremely
sensitive dependence} of $t_{as}$ on the linear dimension of the
cylinder ($t_{as} \propto L^4$) appears to limit the applicability
of our model to values of $L$ up to a few microns. In fact, beyond
this {\it mesoscopic} scale, $t_{as}$ becomes so large as to
render unrealistic the adiabatic piston assumption over this time
interval. As an example, for $\mu = 10^{-2}$ and $L=1$ cm, we
obtain $t_{as} \cong 5 \cdot 10^{12}$ sec, that is about 1000
centuries! Conversely, by taking $L=1$ micron, we get the
reasonable value $t_{as}\cong 5 \cdot 10^{-4}$ sec and $\langle
\Delta S \rangle / k \cong -5 \cdot 10^5$. This corresponds a
violation of the second law in the {\it mesoscopic realm}.

 The above approach has allowed us to deal with the
situation $\mu \ll 1$. In order to have an insight into the
behavior of our process in the more general case $\mu \lesssim 1$,
we assume Langevin's approach to be approximately valid also in
this moderately nonlinear regime, and use the
nonlinear Eq.(\ref{eq3}) with the same stochastic acceleration
$a(t)$ worked out in the linear case. After introducing the
dimensionless units $\xi
 =X/L$ and $\tau=t/t_o$, where $t_o=4 \sqrt{2} t_{th}$, Eq.(\ref{eq3}) reads
\begin{equation} \label{eq7}
\ddot{\xi}+ \left( \frac{1}{\sqrt{\xi}}+\frac{1}{\sqrt{1-\xi}} \right) \dot{\xi}- \frac{1}{\mu}
             \left( \frac{1}{\xi}-\frac{1}{1-\xi} \right) \dot{\xi}^2 = \sigma \alpha (\tau),
\end{equation}
%
where the dot stands for derivative with respect to $\tau$,
$a(\tau)$ is a unitary-power white noise process and $\sigma^2 =
(\pi/2\sqrt{2})(\mu/N)$. This last equation can be numerically
integrated by adopting a second-order leap-frog algorithm as the
one developed in \cite{P9}. In particular, we are able to evaluate
the ensemble average (over 1000 realizations of the piston) of the
time evolution of the mean square-root deviation $\langle x^2
\rangle$ of the piston position. This analysis is reported in
Fig.(\ref{f1}) for $\mu=0.5$ and $N=3 \cdot 10^4$.
Its inspection clearly shows that the piston undergoes random
fluctuations around the central position $X=L/2$, which increase
with time up to an asymptotic value of the order of $L/10$. The
ensemble average of the total entropy change $\langle \Delta S
\rangle / k = N(c_p / R) \langle \ln (1 - 4x^2/L^2) \rangle$ of
our isolated system can also be numerically evaluated. In
Fig.(\ref{f1}), $\langle \Delta S \rangle / k$ is reported as a
function of the normalized time $\tau$.
\begin{figure}
\includegraphics[width=6cm]{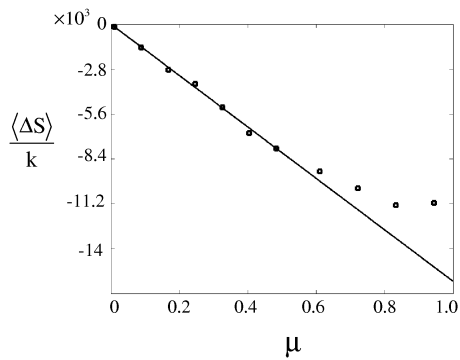}
\caption{Entropy change for $\tau = 10^6$ as a function of $\mu$
and $N = 3 \cdot 10^4$ (over 1000 realizations). The continuous
line represents the predicted behavior in the linear regime.}
\label{f5}
\end{figure}

Further numerical analysis shows (see Fig.(\ref{f5})) that,
keeping $N$ fixed and varying $\mu$, the asymptotic value of
$\langle \Delta S \rangle / k$ undergoes an approximately linear
decrease with $\mu$, up to $\mu=0.6$, as predicted by
Eq.(\ref{eq6}). Beyond this value the relative entropy decrease
slows down, but we are now in a range where Langevin's model may
not be valid. Conversely, in the limit $\mu\rightarrow 0$ our
approach furnishes $(X-<X>)^2/(L/2)^2=\mu/2\rightarrow 0$ and
$t_{as}\rightarrow\infty$, results which are consistent with those
of \cite{Chernov02}.

We wish to note that the celebrated Boltzmann's H-theorem \cite{P10}, which can be considered as a ``proof'' of
the second law, does not apply in our case since one of its main hypotheses, that is the complete
molecular-chaos assumption ({\it Stosszahlansatz}), is not valid in our dynamics: in fact, the correlation
induced by the random motion of the piston favors a common sign of $v_x$ and $v_x'$ in the two-particle
correlation function of the gas near the piston, while the single-particle correlation functions are
independent from the sign of $v_x$ and $v_x'$. This prevents the two-particle correlation distribution to
factorize into the product of the one-particle distribution functions and, thus, the molecular-chaos
assumption does not apply.

Finally, the validity of our conclusions appears to be
corroborated by suitable molecular dynamic simulations of the
evolution of our system. These typically involve a considerable
number of point particles, which model the gas inside the
cylinder, separated by a frictionless piston against which they
undergo perfect elastic collisions. In particular, a microscopic
model consisting of $N=500$ hard disks furnishes a relaxation time
$\tau$ (corresponding to our $t_{as}$) in good qualitative and
quantitative agreement with out results.\cite{Kestemont00}

Besides, numerical investigations of a system consisting of a
number of particles of the order of $10^3$, indicates that the
difference between the temperatures on the two sides undergoes
relevant oscillations, so that the system does not reach
equilibrium, as predicted by our model.\cite{P11}

Further recent molecular dynamic simulations describing the time
evolution of the piston position around $X\cong L/2$ are in fairly
good agreement with our results, both qualitatively and
quantitatively. \cite{White02}

 We wish to thank Noel Corngold for many fruitful
discussions.

\end{document}